\documentclass[aps,twocolumn,showkeys,superscriptaddress]{revtex4}
\usepackage{epsfig}
\usepackage{times}
\begin{document}

\title{Conformity enhances network reciprocity in evolutionary social dilemmas}

\author{Attila Szolnoki}
\email{szolnoki@mfa.kfki.hu}
\affiliation{Institute of Technical Physics and Materials Science, Research Centre for Natural Sciences, Hungarian Academy of Sciences, P.O. Box 49, H-1525 Budapest, Hungary}

\author{Matja{\v z} Perc}
\email{matjaz.perc@uni-mb.si}
\affiliation{Faculty of Natural Sciences and Mathematics, University of Maribor, Koro{\v s}ka cesta 160, SI-2000 Maribor, Slovenia}
\affiliation{Department of Physics, Faculty of Science, King Abdulaziz University, Jeddah, Saudi Arabia}
\affiliation{CAMTP -- Center for Applied Mathematics and Theoretical Physics, University of Maribor, Krekova 2, SI-2000 Maribor, Slovenia}

\begin{abstract}
The pursuit of highest payoffs in evolutionary social dilemmas is risky and sometimes inferior to conformity. Choosing the most common strategy within the interaction range is safer because it ensures that the payoff of an individual will not be much lower than average. Herding instincts and crowd behavior in humans and social animals also compel to conformity on their own right. Motivated by these facts, we here study the impact of conformity on the evolution of cooperation in social dilemmas. We show that an appropriate fraction of conformists within the population introduces an effective surface tension around cooperative clusters and ensures smooth interfaces between different strategy domains. Payoff-driven players brake the symmetry in favor of cooperation and enable an expansion of clusters past the boundaries imposed by traditional network reciprocity. This mechanism works even under the most testing conditions, and it is robust against variations of the interaction network as long as degree-normalized payoffs are applied. Conformity may thus be beneficial for the resolution of social dilemmas.
\end{abstract}

\maketitle

\section*{1. Introduction}
Not only are social interactions limited and thus best described not by well-mixed models but rather by models entailing networks \cite{castellano_rmp09, havlin_pst12, rand_tcs13, helbing_n13, kivela_jcn14}, it is also a fact that these interactions are not always driven by a selfish agenda such as fitness maximization \cite{wilson_00, bentley_11}. Although a high individual fitness, which is most often quantified by a scalar payoff value, is at the heart of success in evolutionary games \cite{maynard_82, weibull_95, hofbauer_98, mestertong_01, nowak_06, sigmund_10}, social interactions are often aimed also at fostering the general sense of belongingness and the identification with a particular group or a way of thinking, with a subculture, or even with a fashionable trend or movement \cite{fiske_09}. And one of the most common actions for achieving this is simply to comply or to conform with the most widespread and established ideas or ideals of the reference group to which one aspires to.

Despite of the fact that payoff maximization is a surprisingly apt description of interactions among simpler forms of life, such as bacteria and plants \cite{griffin_n04, biernaskie_prsb10}, as well as among viruses \cite{turner_n99}, and thus rightfully permeates evolutionary game theory, the consideration of alternative targets, especially for interactions among humans and social animals, appears to be  justified. Of particular relevance in this case are social dilemmas, where the interests of individuals are at odds with what is best for the society as a whole, and none has received as much attention as the prisoner's dilemma game \cite{fudenberg_e86, nowak_n93, imhof_pnas05, fu_epjb07, gomez-gardenes_prl07,cheng_hy_njp11, fu_pre08b, rong_pre10, laird_ijbc12, fu_pre09, antonioni_pone11, dai_ql_njp10, tanimoto_pre12, press_pnas12, sun_jt_njp10, hilbe_pnas13, wu_t_pa14, szolnoki_pre14, rong_epl13, wu_js_pa14}. Each instance of the game is contested by two players who have to decide simultaneously whether they want to cooperate or defect. The dilemma is given by the fact that although mutual cooperation yields the highest collective payoff, a defector will do better if the opponent decides to cooperate. A purely rational payoff-driven player should thus always decide to defect.

Ample research has already been devoted to the identification of mechanisms that may lead to a cooperative resolution of social dilemmas. Classic mechanisms are reviewed in \cite{nowak_s06}, among which network reciprocity due to Nowak and May \cite{nowak_n92b} has motivated an impressive array of studies aimed at understanding the evolution of cooperation in structured populations \cite{szabo_pr07, roca_plr09, perc_bs10, perc_jrsi13, szolnoki_jrsif14}. Methods of statistical physics have proven particularly suitable for this task, as evidenced by the seminal studies of the evolution of cooperation on small-world \cite{abramson_pre01, kim_bj_pre02}, scale-free \cite{santos_prl05, santos_pnas06}, coevolving \cite{ebel_pre02, zimmermann_pre04}, hierarchical \cite{lee_s_prl11}, bipartite \cite{gomez-gardenes_c11}, and most recently also on multilayer networks \cite{wang_z_epl12, wang_z_srep13, wang_z_srep13b, jiang_ll_srep13, szolnoki_njp13}. Moreover, many coevolutionary rules \cite{perc_bs10} have been introduced that may generate favorable interaction networks spontaneously \cite{pacheco_prl06, fu_pre08, zhang_j_pa11, wu_t_pre09, zhang_j_pa10, dai_ql_njp10, lin_yt_pa11, wang_z_njp14}. There is also experimental evidence in favor of the fact that a limited interaction range does play a prominent role by the evolution of cooperation \cite{rand_pnas14, apicella_n12}, especially so if coupled with rewiring \cite{rand_pnas11}.

The key assumption behind existing research, however, has been that every player aspires only to maximizing its own payoff, although this is obviously not always the case. In fact, there exist compelling evidence in favor of the fact that conformity also plays an important role \cite{fiske_09}, especially among humans and social animals. The consideration of conformity dictates the adoption of the strategy that is most common within the interaction range of the player, regardless of the expected payoff \cite{henrich_ehb98, henrich_jtb01}. By adopting the most common strategy, the conformists thus coordinate their behavior in a way that minimizes individual risk and fosters coherence within the population. Unlike previous research, however, we do not assume that this ``cultural transmission'' affects all the players in the population uniformly \cite{henrich_jtb01}. Instead, we take into account the fact that players are diverse in their aspirations, and that thus some are keen on maximizing their payoffs while others are simply content by adopting the most common strategy in their neighborhood. Interestingly, such behavior has recently also been observed in the realm of an economic experiment involving the public goods game with institutionalized incentives \cite{wu_jj_srep14}.

The question that we therefore wish to address in the continuation is: What is the impact of conformity-driven players on the evolution of cooperation in evolutionary social dilemmas? In particular also, what role does the fraction of conformity-driven players within a population play? One may expect that conformity-driven players will push the system towards neutral evolution, especially when they represent the majority of the population. But interestingly, this is not always the case. In what follows, we will show that the introduction of conformists to the prisoner's dilemma game enhances network reciprocity, and thus favors the evolution of cooperation. In particular, we will demonstrate how conformity-driven players introduce spontaneous flocking of cooperators into compact clusters with smooth interfaces separating them from defectors. Furthermore, we will elaborate on the responsible microscopic mechanisms, and we will also test the robustness of our observations. Taken together, we will provide firm evidence in support of conformity-enhanced network reciprocity and show how conformists may be beneficial for the resolution of social dilemmas. First, however, we proceed with presenting the details of the mathematical model.

\section*{2. Evolutionary games with conformists}
We study evolutionary social dilemmas on the square lattice and the Barab\'asi-Albert scale-free network, each with an average degree $k=4$ and size $N$. For the generation of the scale-free network, we implement the standard growth and preferential attachment algorithm \cite{barabasi_s99}. Accordingly, starting from a small number of vertices ($m_0=3$), a new node with $m=2$ edges is connected to an existing node $x$ with probability $\Pi(k_x)=k_x /\sum_y k_y$, where $k_x$ denotes the degree of node $x$. This growth and preferential attachment scheme yields a network with an average degree $k_{av}=2 m$, and a power-law degree distribution with the slope of the line equaling $\approx -3$ on a double-logarithmic scale. The two considered networks are representative for the simplest homogeneous and the strongly heterogeneous interaction topology.

Each player is initially designated either as cooperator ($C$) or defector ($D$) with equal probability, and each instance of the game involves a pairwise interaction where mutual cooperation yields the reward $R$, mutual defection leads to punishment $P$, and the mixed choice gives the cooperator the sucker's payoff $S$ and the defector the temptation $T$. We predominantly consider the weak prisoner's dilemma, such that $T>1$, $R=1$ and $P=S=0$, but we also consider the true prisoner's dilemma in the form of the donation game, where $T=1+b$, $R=1$, $P=0$ and $S=-b$.

We simulate the evolutionary process in accordance with the standard Monte Carlo simulation procedure comprising the following elementary steps. First, according to the random sequential update protocol, a randomly selected player $x$ acquires its payoff $\Pi_x$ by playing the game with all its neighbors. Next, player $x$ randomly chooses one neighbor $y$, who then also acquires its payoff $\Pi_y$ in the same way as previously player $x$. Once both players acquire their payoffs, then player $x$ adopts the strategy $s_y$ from player $y$ with a probability determined by the Fermi function
\begin{equation}
\Gamma(\Pi_x - \Pi_y)=\frac{1}{1+\exp((\Pi_x-\Pi_y)/K)}\,\,,
\label{fermi}
\end{equation}
where $K=0.1$ quantifies the uncertainty related to the strategy adoption process \cite{blume_l_geb93, szabo_pr07}. In agreement with previous works, the selected value ensures that strategies of better-performing players are readily adopted by their neighbors, although adopting the strategy of a player that performs worse is also possible \cite{perc_pre08b, szolnoki_njp08}. This accounts for imperfect information, errors in the evaluation of the opponent, and similar unpredictable factors.

To introduce conformity, we designate a fraction $\rho$ of the population as being conformity-driven, and this influences the strategy adoption rule. In particular, each conformists $x$ simply prefers to adopt the strategy that is most common within its interaction range. Equation~\ref{fermi} thus no longer applies. Instead, if player $x$ is a conformist, we use
\begin{equation}
\Gamma(N_{s_x} - k_h)= \frac{1}{1+\exp((N_{s_x}-k_h)/K)}\,\,,
\label{conform}
\end{equation}
where $N_{s_x}$ is the number of players adopting strategy $s_x$ within the interaction range of player $x$, while $k_h$ is one half of the degree of player $x$. It is worth pointing out that the application of Eq.~\ref{conform} results in the conformity-driven player adopting, with a very high probability, whichever strategy (either $C$ or $D$) is at the time the most common in its neighborhood. Nevertheless, it is still possible, yet very unlikely that a conformist will adopt the strategy that is in the minority. If, however, the number of cooperators and defectors in the neighborhood is equal, the conformity-driven player will change its strategy with probability $1/2$.

In terms of the simulation procedure, we note that each full Monte Carlo step (MCS) consists of $N$ elementary steps described above, which are repeated consecutively, thus giving a chance to every player to change its strategy once on average. All simulation results are obtained on networks typically comprising $N=10^4-10^5$ players, although the usage of larger networks is necessary in the proximity to phase transition points. We determine the fraction of cooperators $f_C$ in the stationary state after a sufficiently long relaxation time lasting up to $10^5$ MCS. To further improve accuracy, the final results are averaged over $400$ independent realizations, including the generation of the scale-free networks and random initial strategy distributions, for each set of parameter values.

\section*{3. Results}
Before presenting the results in structured populations, we summarize briefly the results in well-mixed populations. In the prisoner's dilemma game defectors dominate completely in the absence of conformity-driven players. But if all players are conformists, then everybody looses interest in payoffs and the evolution is simply a random drift. Consequently, the system may terminate into an all$-C$ or an all$-D$ phase, ultimately yielding an average fraction of cooperators $f_C=0.5$. The intermediate region is more interesting, where two different cases have to be considered separately. If $\rho<0.5$, the majority of players is still driven by payoffs, and the system thus eventually terminates into an all$-D$ phase. If $\rho>0.5$, however, the majority of players is conformity-driven, meaning that either an all$-C$ or an all$-D$ phase will ultimately emerge among them, thus giving rise to $f_C=0.5 \cdot \rho$. In sum, in the low $\rho$ region defectors always dominate, while in the high $\rho$ region the average cooperation level can never exceed 0.5, and this regardless of the value of $T$.

\begin{figure}
\centerline{\epsfig{file=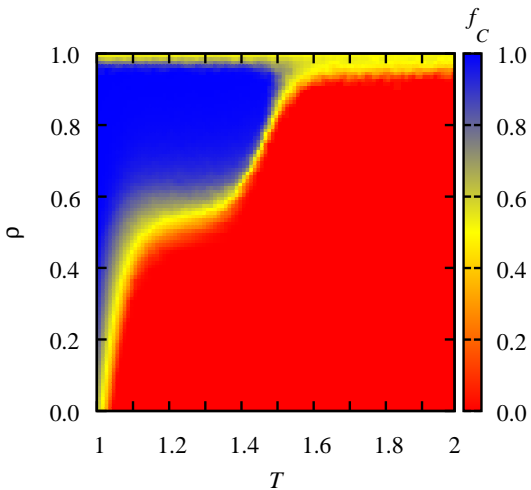,width=7cm}}
\caption{Evolution of cooperation in the weak prisoner's dilemma with conformity-driven players, as obtained on the square lattice in dependence on the temptation to defect $T$ and the density of conformists $\rho$. The color map encodes the stationary fraction of cooperators $f_C$. While, expectedly, $f_C$ decreases with increasing $T$ values, it can also be observed that the dependence of $f_C$ on $\rho$ is nonmonotonous, especially for intermediate values of $T$. The bell-shaped outlay of $f_C$ on $\rho$ is due to conformity-enhanced network reciprocity as $\rho>0$ on the one hand, and the strategy-neutral relation of conformists in the absence of payoff-driven players at $\rho=1$ on the other hand.}
\label{weakpd}
\end{figure}

In structured populations, the presence of conformity-driven players has significantly more unexpected and even counterintuitive consequences. We begin by presenting results obtained with the weak prisoner's dilemma on the square lattice. The color map presented in Fig.~\ref{weakpd} encodes the stationary fraction of cooperators $f_C$ in dependence on the temptation to defect $T$ and the fraction of conformity-driven players within the population $\rho$. It can be observed that the introduction of conformists is able to sustain cooperative behavior at values of $T$ that are well beyond those reachable with traditional network reciprocity alone. More specifically, if the value of $\rho$ is sufficiently large, cooperators are able to dominate in the population up to $T \approx 1.5$. In comparison, when conformists are absent, at $\rho=0$, the maximally attainable level of cooperation is only $f_C=0.64$ at $T=1$, and moreover, defectors dominate completely above $T=1.037$.

A closer look at the results presented in Fig.~\ref{weakpd} reveals also that too many conformity-driven players could impair the evolution of cooperation because among them the evolution of strategies becomes neutral. Accordingly, at $\rho=1$ the population will terminate into a homogeneous all $C$ or all $D$ state with equal probability, thus yielding $f_C=0.5$ on average. Together with the upward trend in $f_C$ as $\rho$ increases above zero, the neutral strategy evolution at $\rho=1$ gives rise to a bell-shaped, nonmonotonous dependence of $f_C$ on $\rho$, which is particularly pronounced at intermediate values of $T$. Based on the results presented in Fig.~\ref{weakpd}, we may thus conclude that it is beneficial for the whole society if the majority of the population consists of conformity-driven players. Nevertheless, a certain fraction of payoff-driven players is necessary to induce symmetry-breaking along the interfaces that separate competing strategy domains. The role of conformists is hence simply to homogenize the population locally, while the role of players seeking to maximize their payoffs is to reveal the long-term benefits of cooperation and thereby to guide the expansion of clusters in the socially desirable direction.

\begin{figure}
\centerline{\epsfig{file=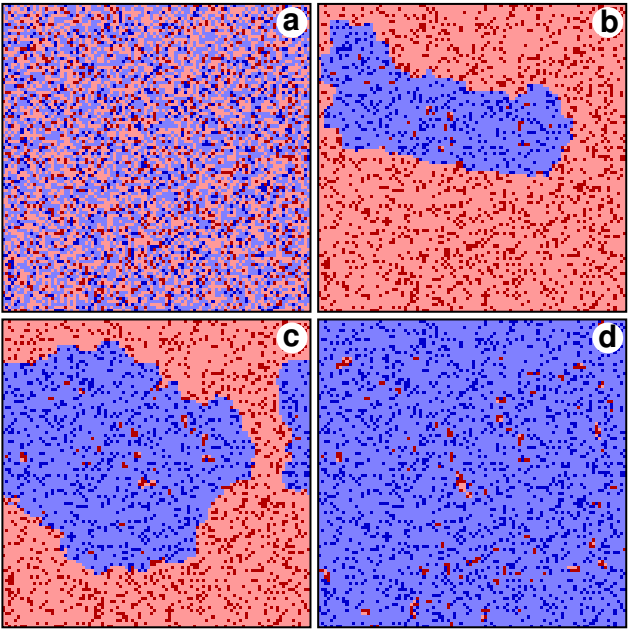,width=8.5cm}}
\caption{Evolution of cooperation from a random initial state under the influence of conformity. Depicted are characteristic spatial patterns, as obtained with the weak prisoner's dilemma game on a square lattice using $T=1.45$ and $\rho=0.81$. Payoff-driven cooperators (defectors) are depicted dark blue (dark red), while conformity-driven cooperators (defectors) are depicted bright blue (pale red). Starting from a random initial state (panel a), conformity-driven players introduce spontaneous flocking of cooperators into compact clusters with smooth interfaces separating them from defectors (panel b). Payoff-driven players subsequently reveal the long-term benefits of cooperation and the cluster grows, all the while maintaining surface tension and thus a smooth interface (panel c). The effectiveness of this conformity-enhanced network reciprocity eventually propels cooperators to near-complete dominance (panel d). For clarity, we have here used a small square lattice with linear size $L=100$.}
\label{snaps}
\end{figure}

To illustrate the microscopic dynamics behind the described conformity-enhanced network reciprocity, we show in Fig.~\ref{snaps} a series of characteristic strategy distributions that describe the time evolution of the game from a random initial state. For clarity, we use different colors not only for cooperators (blue) and defectors (red), but also for distinguishing conformity-driven (pale) and payoff-driven (dark) players. From the outset, defectors spread very efficiently, and indeed only a small flock of cooperators is able to survive (panel b). The compact cluster protects the cooperators from extinction, and this is in fact a very typical time evolution for an evolutionary game that is contested in a structured population. What distinguishes this cluster from an ordinary cluster that would be due solely to network reciprocity is its smooth interface that separates the competing domains. In fact, the dynamics between conformity-driven players is conceptually similar to the so-called majority-voter model \cite{oliveira_jsp92}. It is easy to see that the larger the fraction of conformity-driven players the smoother the interface between the competing domains. Put differently, the propensity of players to comply introduces an effective surface tension around cooperative clusters that is completely absent in traditional voter models, where rough interfaces and slow coarsening are common \cite{dornic_prl01}. Evidently, due to the neutral strategy evolution among conformists, the conformity-induced homogenization alone is just a double-edge sword, but payoff-driven players brake the symmetry in favor of cooperation.

\begin{figure}
\centerline{\epsfig{file=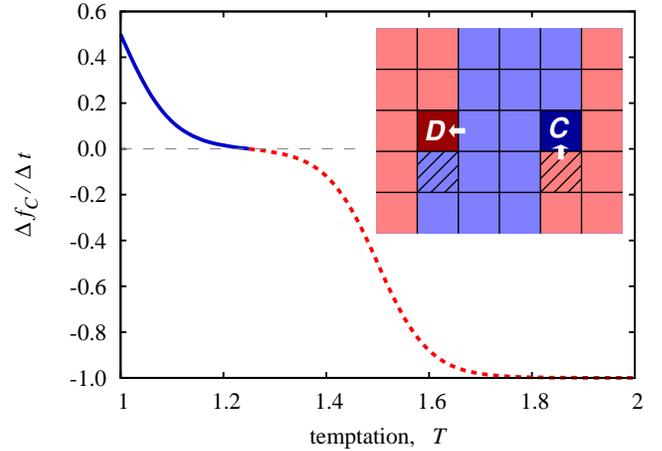,width=8.5cm}}
\caption{The inset features a schematic presentation of two typical interfaces that separate competing domains when payoff-driven players (dark colors) are rare. In the complete absence of such players, conformity-driven players (pale colors) would build perfectly smooth (straight) interfaces. In their presence, however, the interfaces might be modified by the most likely elementary steps, which are marked in the figure as follows: those conformity-driven players who are at the edge of a moving interface (marked by tilted-line boxes) can change their strategy with probability $1/2$, while payoff-driven players are most likely to imitate a strategy along the direction of white arrows. These elementary steps determine the leading terms in Eq.~\ref{dfc}. Main panel shows the time derivative of the fraction of cooperators density in dependence on $T$, according to Eq.~\ref{dfc}, that is due solely to the above-mentioned elementary processes. It can be observed that only for $T > 1.25$ the tide shifts in favor defectors.}
\label{interface}
\end{figure}

When payoff-driven players are rare, the symmetry-breaking due to them can be analyzed in more detail by zooming in on the most likely elementary steps along the interface that separates cooperators and defectors, as shown in the inset of Fig.~\ref{interface}. In the absence of payoff-driven players (dark colors) the most stable interface separating conformity-driven players (pale colors) is a straight line where the most likely change happens at a step. The position of this step propagates randomly until a payoff-driven player is encountered, at which point imitation becomes a possibility. The most probable elementary steps are depicted in the inset of Fig.~\ref{interface}. As noted, conformity-driven players at the step, marked by tilted lines, are able to change their strategy with probability $1/2$. More interestingly, payoff-driven players are able to adopt another strategy, yet they cannot pass their strategy by utilizing their ``success'' because neighboring conformity-driven players do not care about higher payoffs. Accordingly, the most probable invasions are only those marked by white arrows. By summing up these elementary processes, we can estimate the time variation of the fraction of cooperators according to
\begin{equation}
\frac{\Delta f_C}{\Delta t} = \left(\Gamma (2T-3R) - \frac{1}{2} \right)+\left(\frac{1}{2}-\Gamma(2R-2T)\right),
\label{dfc}
\end{equation}
where the $\Gamma$ function is defined as in Eq.~\ref{fermi}. We plot $\frac{\Delta f_C}{\Delta t}$ in dependence on the temptation to defect in Fig.~\ref{interface}. The result suggests that cooperation will spread below a threshold value, even if $T>R$, as a consequence of the broken symmetry described above. Indeed, this simple approximation is able to explain why there is a relatively sharp transition between the full $C$ and the full $D$ evolutionary outcome in the high $\rho$ region (see Fig.~\ref{weakpd}).

\begin{figure}
\centerline{\epsfig{file=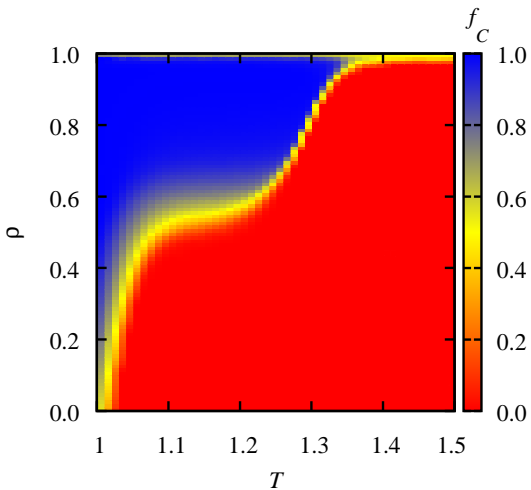,width=7cm}}
\caption{Evolution of cooperation in the donation game (true prisoner's dilemma) with conformity-driven players, as obtained on the square lattice in dependence on the temptation to defect $T$ and the density of conformists $\rho$. The color map encodes the stationary fraction of cooperators $f_C$. Results are qualitatively similar to those presented in Fig.~\ref{weakpd} for the weak prisoner's dilemma game, thereby confirming the robustness of the enhanced network reciprocity to variations in the contested social dilemma. It is also worth noting that network reciprocity alone is practically unable to sustain cooperation in the donation game if $T>1$, which indicates that the identified conformity-enhanced network reciprocity works well even under the most testing conditions.}
\label{donation}
\end{figure}

We proceed by testing the robustness of conformity-enhanced network reciprocity, first by considering an alternative formulation of the social dilemma. So far, to obtain results that are comparable with previous related works \cite{szabo_pre05,cheng_hy_epjb13}, we have focused on the weak prisoner's dilemma, which does not constitute the most adverse conditions for the successful evolution of cooperation since the punishment for mutual defection and the suckers payoff are equal ($P=S=0$). To amend this, we consider the donation game, where the payoff ranking $T>R>P>S$ corresponds to the true prisoner's dilemma. The consideration of this game is all the more interesting because network reciprocity alone is virtually unable to sustain cooperation under such testing circumstances \cite{nowak_n92b}. As for the weak prisoner's dilemma in Fig.~\ref{weakpd}, for the donation game too we present a color map that encodes the stationary fraction of cooperators $f_C$ in dependence on the temptation to defect $T$ and the fraction of conformity-driven players $\rho$ in Fig.~\ref{donation}. It can be observed that, even in the most challenging social dilemma, conformity-enhanced network reciprocity is able to ensure widespread dominance of cooperators at remarkably high values of $T$, as long as the value of $\rho$ is sufficiently large. However, as $\rho \to 1$, the lack of payoff-driven players introduces  the status quo among conformity-driven players, and again the average fraction of cooperators drops to $f_C \approx 0.5$ and thus gives rise to the bell-shaped, nonmonotonous dependence of $f_C$ on $\rho$.

Lastly, we explore the robustness of conformity-enhanced network reciprocity to changes in the topology of the interaction network. Since the square lattice is representative for regular, homogeneous interaction networks, the most interesting test involves considering the highly heterogeneous scale-free network. First, we have to emphasize that the consideration of absolute payoffs on strongly heterogeneous networks already provides ample support to network reciprocity, in particular by ensuring homogeneous strategy ``clouds'' around hubs \cite{santos_prl05, gomez-gardenes_prl07}. The introduction of conformity-driven players is therefore either negligible or even negative. Namely, if a defective hub is designated as conformity-driven, then it is almost impossible to revert its strategy to cooperation because the large number of like-minded defective followers will always assure the hub it already conforms perfectly with the neighborhood. Since previous research has highlighted the main cooperator-promoting mechanism on scale-free networks is based on the fact that defective hubs eventually become weak and thus vulnerable to strategy change \cite{szabo_pr07}, the introduction of conformists can be a notable drawback because this key mechanism becomes disabled.

\begin{figure}
\centerline{\epsfig{file=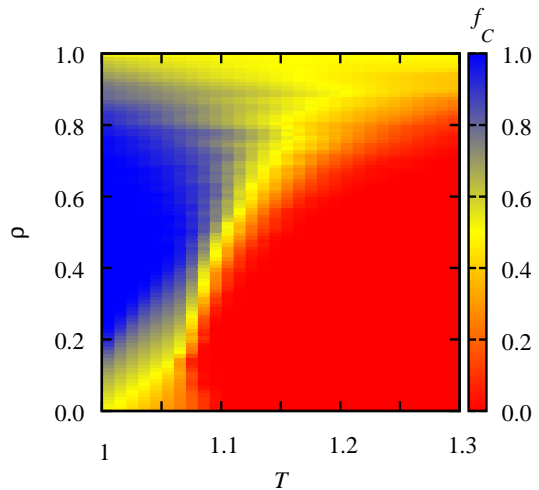,width=7cm}}
\caption{Evolution of cooperation in the weak prisoner's dilemma with conformity-driven players, as obtained on the scale-free network in dependence on the temptation to defect $T$ and the density of conformists $\rho$. The color map encodes the stationary fraction of cooperators $f_C$. The change in the topology of the interaction network also leaves the results qualitatively unaffected, thus further corroborating the robustness of the identified conformity-enhanced network reciprocity.
Importantly, we have here applied degree-normalized payoffs. If absolute payoffs are applied on strongly heterogeneous networks, then the heterogeneity alone provides the maximal support to network reciprocity, and hence the impact of conformity is either negligible or even slightly negative (not shown).}
\label{scalef}
\end{figure}

However, since the application of absolute, cumulative payoffs on strongly heterogeneous interaction networks already raised questions during the early stages of research on this subject \cite{santos_jeb06, masuda_prsb07, szolnoki_pa08}, in particular in the sense that a player might be unable to maintain a large number of connections for free, the application of degree-normalized payoffs was proposed. It was shown that the application of such payoffs erases the ability of heterogeneous networks to sustain large homogenous cooperative clusters around hubs, and that the cooperation levels return to those observed earlier on regular networks and lattices. With this, we arrive yet again at conditions where the presence of conformity-driven players might help cooperation significantly, and indeed the results presented in Fig.~\ref{scalef} fully confirm this expectation. All qualitative features remain the same as by the consideration of the weak prisoner's dilemma and the donation game on the square lattice (compare with Figs.~\ref{weakpd} and \ref{donation}), and with this we conclude that conformity-enhanced network reciprocity is also robust against variations of the interaction network as long as degree-normalized payoffs are applied.

\section*{4. Discussion}
We have studied the evolution of cooperation in social dilemmas where a fraction of the population has been designated as being driven by conformity rather than payoff maximization. Unlike in traditional evolutionary game theory, these conformists are no longer concerned with maximizing their payoffs by selecting the most promising strategy for future interactions. Instead, conformity-driven players simply adopt whichever strategy is most common within their interaction range at any given time. We have shown that the presence of conformity-driven players enhances network reciprocity, and thus aids the favorable resolution of social dilemmas. The effectiveness of conformity to do so, however, depends on the fraction of conformists within the population. If the later are neither too rare nor too common, the flocking of cooperators into compact clusters with smooth interfaces emerges spontaneously. But since the strategy preference among conformists is neutral, a certain fraction of payoff-driven players is necessary to induce symmetry-breaking along the interfaces that separate competing domains. Based on this, we have demonstrated that cooperative clusters are able to expand past the boundaries imposed by traditional network reciprocity, in particular because the dynamics between conformity-driven players is conceptually similar to the so-called majority-voter model \cite{oliveira_jsp92}. We have emphasized that the larger the fraction of conformity-driven players, the smoother the interface between the competing domains. In other words, the tendency of conformist to blend in introduces an effective surface tension that does not exist in traditional voter model where rough interface and slow coarsening can be observed \cite{dornic_prl01}.

We have shown that the newly identified mechanism is robust to variations of the contested social dilemma, and that it works even under the most testing conditions where traditional network reciprocity completely fails to sustain cooperative behavior. Moreover, we have shown that conformity promotes cooperation regardless of the properties of the interaction network, as long as degree-normalized payoffs are applied. If absolute payoffs are applied on strongly heterogeneous networks, then the heterogeneity alone provides the maximal support to network reciprocity, and hence the impact of conformity-driven players is either negligible or even negative. If namely by chance a defective hub is designated as conformity-driven, then it is almost impossible to revert its strategy to cooperation, and this thus disables the key mechanism that ensures elevated levels of cooperation in heterogeneous networks \cite{szabo_pr07}.

The presented research has been motivated by the fact that payoff maximization alone is often not the primary goal of social interactions. Unlike interactions among simpler forms of life, interactions among humans and social animals are often driven by a desire to belong or to ``fit in'' \cite{fiske_09}. To conform is thereby simply a frequently used way for achieving this goal, and it is interesting and to a degree counterintuitive to discover that conformity may actually promote the evolution of cooperation. Our results of course take nothing away from payoff maximization as an apt and in fact comprehensive motivator of interactions among bacteria, plants and viruses, but they do suggest that conformity might had an evolutionary origin in as far as it furthers prosocial behavior. Furthermore, in addition to possible emotional origins of being a conformist, there might be cases when this preference is actually a payoff-maximizing strategy (by avoiding punishment, for instance). Hence conformism may be a good heuristic in social decision making \cite{rand_n12, rand_nc14}.

We conclude by noting that our model is just an initial step towards the introduction of ``multi-target'' evolutionary games, which ought to properly take into account the diversity of not only the applied strategies, but also the diversity of individual targets one may hope to achieve by adopting them. An interesting direction for future research might involve players being able to change their motivation over time, for example through ``cultural transmission'' \cite{henrich_ehb98}. Relative times scales in evolutionary dynamics \cite{szolnoki_epjb09} could also play an important role, in the sense that the typical time for ``motivational change'' might be different from the typical time in which players change their strategy. To explore the consequences of these options appears to be an exciting venture with many relevant implications, and we hope that this paper will motivate further research along this line in the near future.

\begin{acknowledgments}
This research was supported by the Hungarian National Research Fund (Grant K-101490), the Slovenian Research Agency (Grant P5-0027), and by the Deanship of Scientific Research, King Abdulaziz University (Grant 76-130-35-HiCi).
\end{acknowledgments}

\end{document}